\documentclass[12pt]{article}
\usepackage{a4,epsfig}
\begin{document}
\input epsf

\begin{titlepage}
~\vspace{3cm}
\begin{center}
\vskip2cm
{\Large\bf Analytic continuation  and perturbative expansions in QCD}\\
\vspace{1.0cm}
Irinel Caprini\\
National Institute of Physics and Nuclear Engineering, POB MG 6, \\ Bucharest,
R-76900 Romania, and \\ The Abdus Salam International Centre for Theoretical
Physics, Trieste, Italy \\ 
\vskip0.3cm
Jan Fischer\\
Institute of Physics, Academy of Sciences of the Czech Republic, \\
CZ-182 21  Prague 8, Czech Republic

\end{center}
\vspace{1.2cm}
\begin{abstract} Starting from the divergence pattern of perturbative 
quantum chromodynamics, we propose a novel, non-power series replacing the 
standard expansion in powers of the renormalized coupling constant $a$. The
coefficients of the new expansion are calculable at each finite order from the
Feynman diagrams, while the expansion
functions, denoted as $W_n(a)$, are defined by analytic continuation
in the Borel complex plane.  The infrared ambiguity of
perturbation theory is manifest in the prescription dependence of the 
$W_n(a)$. We prove that the  functions 
$W_n(a)$ have branch point and essential singularities at the origin $a=0$ of
the complex $a$-plane and their perturbative expansions in powers of $a$ are 
divergent, while the expansion of the correlators in terms of 
the $W_n(a)$ set is convergent under quite      
loose conditions. \end{abstract}  \end{titlepage}

 \newpage
\section{Introduction} \label{intro}
The standard QCD perturbative expansion in powers of the renormalization group
improved coupling  is plagued by several
deficiencies. The series is divergent and even Borel non-summable, having a
zero convergence radius  \cite{tHooft}-\cite{Khuri1} and coefficients
exhibiting  asymptotically factorial growth and nonalternating signs
\cite{Zakh}-\cite{Broa}.  The truncated, fixed order perturbative expansion
is afflicted with  the problem of renormalization scheme dependence and 
violates explicitly, due to the Landau singularities, the rigorous momentum
plane analyticity imposed by general principles for confined theories \cite{Oehme}.
Several resummations or reformulations of the  perturbative expansion   have
been considered recently, trying to cure or reduce these deficiencies.  
Modified expansions which eliminate the unphysical singularities in the
momentum  plane were proposed in  \cite{Milt}, \cite{Shir}. 
Reformulations of the perturbation theory attempting to reduce  the
renormalization scheme dependence of the truncated series were also  proposed 
by several authors, using concepts like minimal sensitivity \cite{Stev},
effective charge \cite{Grun1}, \cite{Max} or Pad\'e approximants in  the
complex plane of the coupling constant \cite{SaEl}-\cite{Jent}.

Other attempts to go beyond the conventional perturbative expansion are
motivated by the large order behaviour of perturbation theory
\cite{BBB}-\cite{BeBr}. The  perturbation theory is intrinsically ambiguous,
due to the infrared regions of the Feynman diagrams.  According to current
views \cite{Bene1}, the perturbative ambiguity will be compensated in the
complete theory by the nonperturbative contributions. But the estimate of this
intrinsic ambiguity  is made difficult by the fact that the series is
divergent, and the higher terms dramatically spoil the accuracy of the result
above a certain order. If one succeeded to replace the usual perturbative
expansion by a convergent series (each new term improving the accuracy of the
approximation instead of spoiling it), the intrinsic ambiguity of perturbation
theory would be better defined.  In the present work  we address this problem
and propose a new, non-power perturbative expansion in QCD, using the
principle of analytic continuation in the Borel complex  plane. 

Although
the perturbative series in QCD is  not Borel summable, since 
the conditions for Borel summability  \cite{Hardy}, \cite{Fischer} are
not satisfied,  the notion of Borel transform of the QCD correlators is
nevertheless useful, as its singularities in the Borel plane contain much
physical information. 
The infrared (IR) and ultraviolet (UV) renormalons and
the instantons  define a generic  doubly-cut Borel plane,
which  provides an intuitive measure of the ambiguities of the perturbation 
theory.   

The conformal mapping of the Borel plane was suggested in
\cite{Muel} as a technique to reduce or eliminate the ambiguities (power 
corrections) due to the large momenta in the Feynman integrals. Such conformal
mappings, which move the UV renormalons further from
the origin \cite{Alta}, \cite{SoSu} exploit however only in part  the known
singularity structure in the Borel plane. An optimal mapping, which performs 
the  analytic continuation in the whole doubly-cut Borel plane, was considered
in our previous papers \cite{CaFi}, \cite{CaFi1} (similar mappings were
applied  later in \cite{Jent1}, \cite{CvLe}).  By means of this technique, we
defined in \cite{CaFi1} a  non-power perturbative expansion in QCD in terms
of a new set of functions that fully exploit the location of the singularities
in the Borel plane.

 In the present paper we study the properties of these expansion
functions, denoted below as $W_n(a)$, and the role of the modified expansion
for a better understanding of the ambiguities inherent in perturbative QCD. 
The results are impressive: in contrast to the perturbative powers $a^n$, the
new functions $W_n(a)$ provide a convergent expansion of the correlator in a 
large domain of the coupling constant complex plane, and share in addition
certain important properties with the expanded correlator itself. 

The paper is organized as follows:
in section 2 we recall the basic  facts known about the  singularities of
the QCD  correlation functions in the Borel plane and discuss
the method of optimal conformal mapping. We use for illustration the Adler
function in massless QCD, but the generalization to other cases is
straightforward.  In section 3 we define the functions $W_n(a)$  and the 
modified perturbative expansion replacing the standard series in powers of the
coupling constant. We investigate the analytic properties  of the expansion
functions in the complex $a$-plane, their asymptotic expansions for small $a$,
and the convergence  conditions for the new expansion. In the concluding
section 4, we point out the specific merits of the modified expansion  for
defining  the intrinsic ambiguity of 
perturbation theory.

\section{Conformal mapping of the Borel plane}\label{sec:conf}
\subsection{Singularities in the Borel plane}\label{sec:sing}

 We consider the Adler function  in
massless QCD
\begin{equation}\label{Ddef}
D = - 4\pi ^2 s 
\left(\frac{{\rm d}}{{\rm d}s}\right) \Pi (s)  \,,
\end{equation}
where $\Pi (s)$ is the amplitude of the electromagnetic
current--current correlation  function
\begin{eqnarray}
\Pi ^{\mu \nu}(p) = i \int {\rm d}^4 x {\rm e}^{-ip x} 
\langle 0 | \, {\rm T} (j^{\mu}(x) j^{\nu}(0)) \, | 0 \rangle \nonumber\\ 
=(g^{\mu \nu}p^2 - p^{\mu}p^{\nu}) \Pi (s)\,,\quad s=p^2\,.
\end{eqnarray}
In  perturbative QCD, the function $D$ can be formally  expanded in powers
of the renormalized coupling constant $\alpha_s(\mu^2)$ 
  \begin{equation}\label{D} 
D = \sum_{n=0}^{\infty} D_{n}\left({\alpha_s(\mu^2)\over
\pi}\right)^{n} \,, \end{equation} 
where the coefficients $D_n$  depend on the external momentum squared
$s$, and are  renormalization scheme and scale
dependent. Certain classes of Feynman diagrams suggest that they are
factorially increasing with $n$, therefore the series (\ref{D}) is  divergent,
and has to be given a precise meaning. 

Among various summation techniques of
power series \cite{Hardy}, \cite{Fischer}, the Borel method received much
interest in recent years, although the mathematical conditions required for 
its use are not  satisfied in QCD \cite{tHooft}, \cite{Khuri1}. As mentioned 
in the Introduction, even if the Borel summation method is not 
applicable it is useful to examine and exploit the singularity 
structure of the Borel transform,  as it contains important
physical information.  In the present paper we use 	
the Borel transform to define an improved perturbative 
expansion, using in addition the principle of analytic continuation.

The Borel transform of the Adler function is defined by the power series
\begin{equation}\label{B}  B(u, s )=\sum\limits_{n=0}^\infty b_n u^n\,, 
\end{equation}  with  $b_n$ related to the  original
perturbative  coefficients appearing in (\ref{D}) by \begin{equation}\label{bn}
b_n={D_n\over \beta_0^n n!}\,,\end{equation} 
where $\beta_0$ is the first coefficient of the
$\beta$ function. 
 Then  the Adler function given
by the series (\ref{D}) can be formally  expressed as  the  Laplace transform
\begin{equation}\label{Laplace}  D(s, a)={1\over a
}\,\int\limits_0^\infty\!{\rm e}^{-u/a} \, B(u, s)\,{\rm d}u\,,\end{equation}
where $a=\beta_0 \alpha_s(\mu^2)/\pi$.  

The Borel
transform $B(u,s)$  is scheme and scale dependent. As discussed in
\cite{Bene1}, in the large $N_f$ limit  the momentum dependence  can be
factorized as \begin{equation}\label{factoriz} B(u, s)= \left({-s {\rm e}^C
\over \mu^2}\right)^{-u} B_0(u)\,, \end{equation}  where $C$ is a
scheme-dependent constant. The power factor in (\ref{factoriz}) combines with
the exponential in (\ref{Laplace}) in a renormalization scheme and scale
independent quantity, the remaining factor $B_0(u)$ also being scheme and
scale independent.  The validity of a similar factorization beyond the  large
$N_f$ limit is an  open question \cite{Bene1}. However, it is generally
assumed that the position of the singularities of the function $B(u,s)$ in the
Borel $u$ plane  is independent of renormalization scheme and scale, and also
of the external momenta (as discussed in \cite{Bene1}, this is expected to be
the case at least in the so-called "regular schemes"). 

\begin{figure}\centering
\epsfig{file=  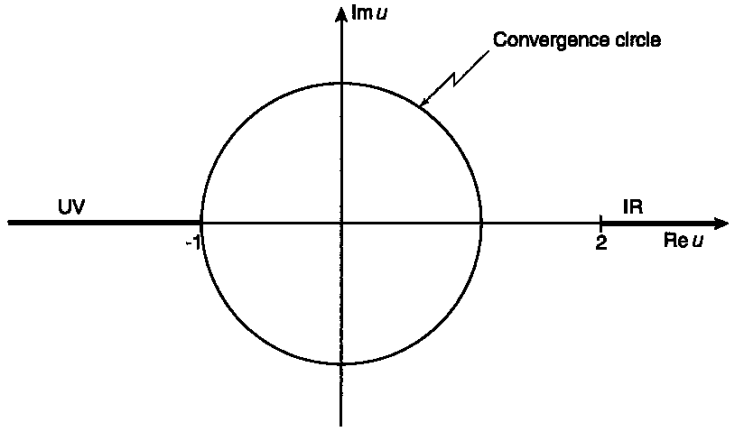,width=10cm}
\caption{ The Borel plane for the Adler function.}
\label{fig:1}
\end{figure}

The method
developed in the present paper is based on the analytic continuation in the
Borel plane, requiring therefore the knowledge of the singularity structure
in this plane.  We shall assume  that the location of the singularities of the
Borel transform  is known and is independent of the renormalization scheme.
For the Adler function,  the coefficients $D_n$   are assumed to have the
specific  large order increase   \begin{equation}\label{Dn}  D_n \sim
\sum\limits_k C_k\, n!\, n^{\delta_k} \left({\pi \beta_0\over k}\right)^n\,, 
 \end{equation}  where the 	
 $k$ in the sum are integers, with  $C_k$ and $n_k$
depending in general on $s$ and $\mu^2$.  This behaviour leads  to  branch
point  singularities for the function $B$ in the $u$-plane, along the negative
axis (the ultraviolet renormalons) and the positive axis (the  infrared
renormalons). Specifically, the branch cuts are situated  along the rays $ u
\leq -1$ and $u \geq 2$ (see Fig. \ref{fig:1}), and the nature of the first
branch points  was established in \cite{Muel} and in \cite{BBK}. The additional
singularities due to the instanton--anti-instanton pairs \cite
{Lipa} are situated at larger positive values of $u$, and will not influence
the method discussed in this paper.

\subsection{Optimal conformal mapping}\label{sec:opt}
\begin{figure}\centering
\epsfig{file=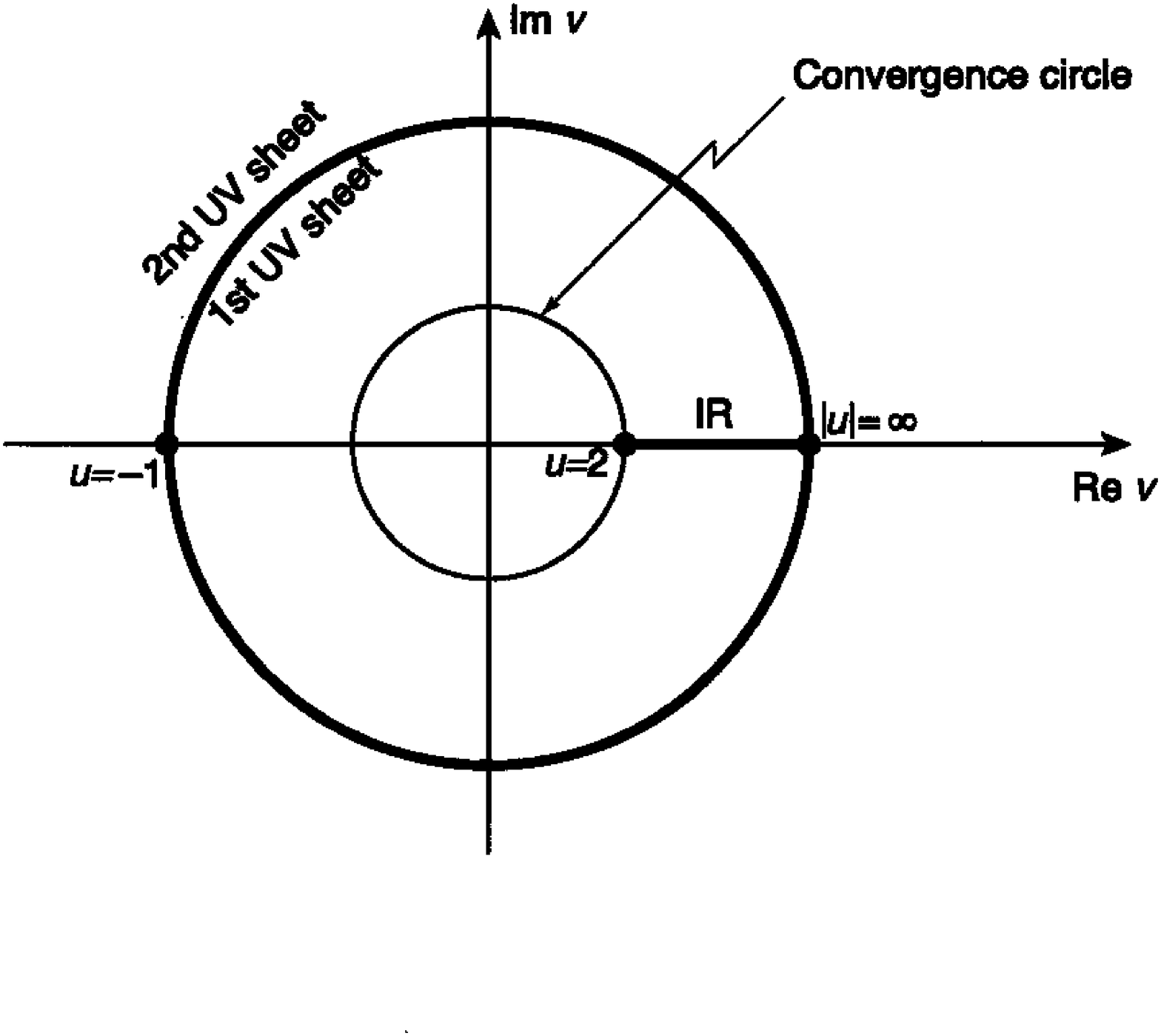,width=8cm}
\caption{The $v$-complex plane. The UV cut is mapped upon the unit
circle, while the IR cut is situated inside it. The convergence domain of the
series (\ref{Bv}) is limited by 
$u=2$, the lowest branch point of the IR cuts family.}
\label{fig:2r}
\end{figure}

The singularities of the Borel transform produce the perturbative
ambiguities of the QCD correlators. For the Adler function, the
Taylor expansion (\ref{B}) of the Borel transform converges on the disk
$|u|<1$ that reaches the nearest singularity, which is the first UV
renormalon. The corresponding  power correction ambiguity  is  however related
 to  large  momenta in the Feynman diagrams, and  can be minimized or
eliminated in QCD within perturbation theory. As suggested in Ref.
\cite{Muel}, this can be achieved by an analytic continuation outside the
convergence circle, by means of a conformal mapping.  This method was
investigated in \cite{Alta}, \cite{SoSu}, using the mapping 
\begin{equation} v(u) = \frac{\sqrt{1+u}-1}{\sqrt{1+u}+1} \,, \label{v}
\end{equation}
and the corresponding expansion \begin{equation}
B(u,s)=\sum\limits_{n=0}^\infty d_n v^n(u)\,,
\label{Bv}
\end{equation} instead of the power series  (\ref{B}).
The mapping (\ref{v}) transforms the UV cut $u \leq -1$ in the
$u$-plane onto the unit circle $|v|=1$ in the $v$ plane, while the IR cut
$u \geq 2$ remains inside the disk $|v|<1$ (see Fig. \ref{fig:2r}).
As a consequence, the series (\ref{Bv}) converges inside the small circle, 
which touches the image of the lowest IR renormalon in the $v$-plane. Thus, 
although the UV ambiguities (power corrections) are in principle eliminated  
by this method,   the series (\ref{Bv}) remains divergent along the positive 
semiaxis,  the IR cut being outside the convergence circle.

\begin{figure}\centering
\epsfig{file=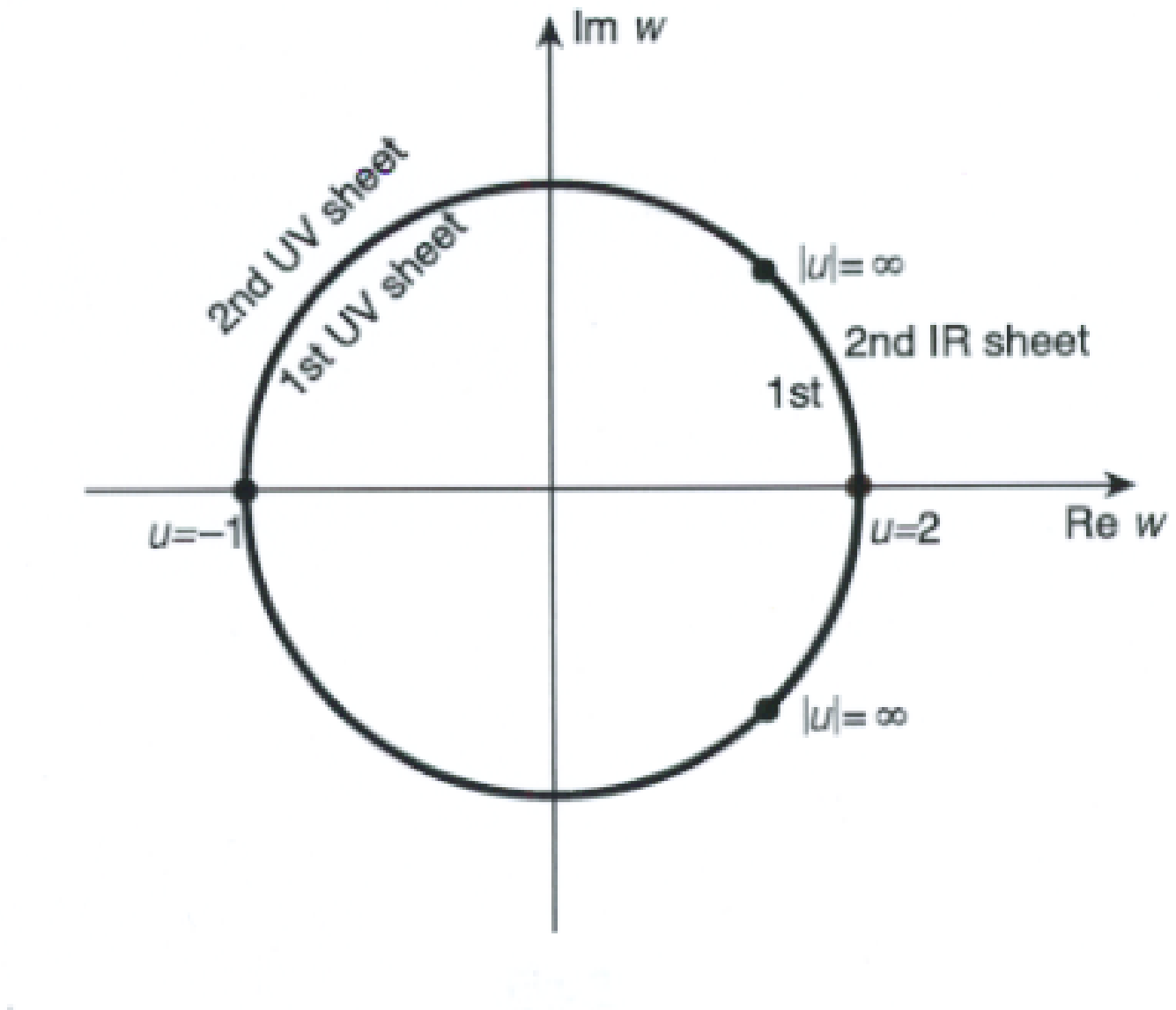,width=7.7cm}
\caption{The $w$-complex plane. Both the UV cut and the IR cut are mapped
upon the boundary unit circle. Inside it, there is no singularity. The
convergence domain is the whole (cut) Borel plane.}
\label{fig:3r}
\end{figure}

The analytic continuation of 
the perturbative Borel transform into the whole holomorphy domain ${\cal D}$
can be performed by an  optimal conformal mapping, $w(u)$, 
which maps the whole ${\cal D}$ onto the unit disk \cite{CiFi}. For the
Adler function this mapping is
\cite{CaFi} \begin{equation}
w(u) = \frac{\sqrt{1+u}-\sqrt{1-u/2}}{\sqrt{1+u}+\sqrt{1-u/2}}
\,. \label{w1} \end{equation}
 The two branch points of $w(u)$, $u=-1$ and $u=2$,
coincide with the lowest branch points of the Borel transform $B(u,s)$, while
the corresponding two cuts, $u \leq -1$ and $u \geq 2$, cover the other branch points of
$B$. As shown in Fig. \ref{fig:3r}, these two rays are mapped by (\ref{w1})
onto the boundary circle of the unit disk  $|w|=1$ in the $w-$plane. The
 inverse of the mapping (\ref{w1}) is
\begin{equation}\label{uw}
u={8 w\over 3 w^2-2 w +3}= {8 w\over 3 (w-\zeta) (w-\zeta^*)}\,,
 \end{equation}
where $\zeta= (\sqrt{2}+i)/(\sqrt{2}-i)$ and its complex conjugate  $\zeta^*$
are the images of $u=\infty$ on the unit circle.

As proposed in \cite{CaFi}, we expand $B$  in powers of the variable $w$
\begin{equation}
B(u, s) = \sum_{n=0}^{\infty} c_{n} w^n(u) ,
\label{Bw}
\end{equation}
where the coefficients $c_{n}$ can be obtained from  the coefficients $b_{k}$,
$k\leq n$, using Eqs. (\ref{B}) and  (\ref{w1}).  
By expanding $B(u,s)$ according to (\ref{Bw}) one makes full use of its
holomorphy domain, because the known part of it
({\em i.e.} the first Riemannian sheet) is
mapped onto the convergence  disk\footnote{The conformal mapping used here is
confined to the first sheet. If the branch point structure
and analyticity on other sheets were known, a multi-sheet mapping could be used
which simultaneously uniformizes several branch points  \cite{CCF}.}. The
series (\ref{Bw}) converges inside the whole disk $|w|<1$, {\em i.e.,} in the
whole cut $u$ plane, up to the cuts. A very important
property, proved in \cite{CiFi}, is that the  expansion  (\ref{Bw})  has  the
fastest asymptotic (large-order) convergence rate, compared to any other
expansion in powers of a variable that maps only a smaller part of the 
holomorphy domain onto the unit disk. We recall		
that the large-order convergence rate of a power series is equal to that of 
the geometrical series with the quotient $r/R$, 
$r$ being the distance of the point from the origin, and $R$ the convergence 
radius. 
The  proof given in \cite{CiFi} consists in comparing the magnitudes of the 
ratio  $r/R$ 
for a certain point in different complex planes, corresponding to different
conformal mappings. When the whole analyticity domain ${\cal D}$  of the
function is mapped on a disk, the value of $r/R$ is minimal \cite{CiFi}.

\section{Modified QCD perturbative expansion} \label{sec:modif}
\subsection{Non-power expansion functions in QCD}\label{sec:nonpower}

The expansion (\ref{Bw}) of the Borel transform suggests  to expand  the Adler
function in the series \cite{CaFi, CaFi1}   \begin{equation}
D(s, a)= \sum_{n=0}^{\infty} c_{n} W_{n}(a)\, , 
\label{cW}
\end{equation}
where the functions $W_n(a)$ are defined as Laplace transforms of $w^n (u)$ :
\begin{equation}\label{Wn}
W_{n}(a)={1\over a}\int\limits_0^\infty\, {\rm e}^{-u/a}\, 
w^n(u)\,{\rm d}u\,,\quad n=0,1,2, \ldots \,.
\end{equation}
At each finite truncation order $N$, the expansion (\ref{cW}) is obtained
 by inserting the series (\ref{Bw})  into the Laplace integral (\ref{Laplace})
and exchanging the order of summation and integration. This procedure is
trivially allowed at any finite integer $N \geq 0$. For $N\to\infty$, however, 
the new expansion  (\ref{cW}) represents  a nontrivial step out of 
perturbation theory, replacing  the perturbative powers
$a^n$ by the functions $W_{n}(a)$. As proved in \cite{CaFi1},  the
expansion (\ref{cW}) converges under certain rather general conditions
(we discuss this point below in subsection \ref{sec:large}). The 
convergence of the series (\ref{cW}) is 
our key argument in favor of the stepping out of the standard perturbation 
theory.

Our procedure is an obvious generalization of the conformal mapping method 
proposed in \cite{Zinn}  
for Borel summable functions.  Formally, the expansion (\ref{cW}) is obtained
from the standard  perturbative expansion (\ref{D}) by replacing the
coefficients $b_n$, appearing in the Taylor series (\ref{B}), by the
coefficients $c_n$ of the improved expansion (\ref{Bw}), and the
perturbative functions  $n! a^n$ (which multiply  the coefficients
$b_n$) by the new functions $W_{n}(a)$ defined  by the integral (\ref{Wn}).  
Actually,  this integral is not well-defined since the branch-point $u=2$ is
encountered along the integration range. This is a manifestation of the
intrinsic ambiguity of the perturbation theory produced by the infrared
renormalons.  In  defining the functions $W_n$, we shall use the same
prescription as the one adopted in (\ref{Laplace}) for the correlator $D$
itself.  We shall consider in particular the functions
\begin{equation}\label{pm} W_{n}^{(\pm)}(a)={1\over a}\int\limits_{\cal
C_\pm}\, {\rm e}^{-u/a}\,  w^n(u)\,{\rm d}u\,,\quad n=0,1,2, \ldots \,,
\end{equation} where ${\cal C_+}$ (${\cal C_-}$) are lines parallel  to the
real positive axis, slightly above (below) it,   and the  principal value (PV)
prescription \begin{equation}\label{pv} W_{n}^{(PV)}(a)={1\over 2
}[W_{n}^{(+)} (a) + W_{n}^{(-)} (a)]\,. \end{equation}  In what follows we
shall examine the  expansion functions $W_n(a)$, showing that in many respects
they resemble the expanded function  $D(s, a)$ itself.

\subsection{Analytic properties of $W_{n}(a)$ in the complex $a$-plane}
\label{sec:analyt}
The problem of the analytic properties of the QCD correlators in the coupling
constant plane is very complex. 't Hooft \cite{tHooft} and
Khuri \cite{Khuri1} showed that renormalization group invariance and the
multiparticle branch-points on the timelike axis of the momentum plane $s=p^2$
 imply a complicated
accumulation of singularities near the point $a=0$. Since the proof uses a
nonperturbative argument (multiparticle states generated by confinement in
massless QCD),  it is difficult to see this feature in perturbation theory,
even in partial resummations that take into account infinite classes of
Feynman diagrams.

 In the present subsection we shall discuss the
analytic properties of the expansion functions  $W_{n}(a)$ defined by the
integrals (\ref{pm}).  Since the integrand is bounded, $|w^n(u)|<1$, the integrals
 converge for  ${\rm Re} a >0$, {\em i.e.} $-\pi/2+\delta \le \psi
\le \pi/2-\delta$, for $\delta >0$ arbitrarily small, where $\psi$ is the
phase of $a$ ($a=|a| {\rm e}^{i \psi}$).  Let us  consider the functions 
$W_{n}^{(+)}(a)$  and take first $a$
complex in the first quadrant, {\em i.e.} $0<\psi < \pi/2$.  Then we
can rotate  the integration contour ${\cal C}_+$ in  (\ref{pm}) towards the
upper quadrant by an angle $\psi$, since the integrand is analytic. More
exactly, we  apply the Cauchy theorem for a closed contour going along
${\cal C}_+$ up to $|u|=R$, continued with the sector of the circle $u= R {\rm
e}^{i \phi}$,  with $0<\phi < \psi$, and the segment $u=\tau {\rm e}^{i
\psi}$, with  $0<\tau <R$. This gives \begin{eqnarray}\label{contour}
&&\frac{1}{a}\int\limits_{0}^{R} {\rm e}^{-u/a}\,w^n(u)\,{\rm d}u + {R \,i
\over a} \int\limits_0^\psi {\rm e}^{i\phi} \,\exp \left[{-R {\rm
e}^{i(\phi-\psi)}\over |a|}\right]\, w^n( R{\rm e}^{i\phi})\, {\rm d} \phi=
\nonumber\\ &&{ {\rm e}^{i\psi} \over a}  \int\limits_0^R\, {\rm e}^{-\tau
/|a|}\, w^n(\tau {\rm e}^{i\psi})\,{\rm d} \tau = \int\limits_0^{R/|a|} {\rm
e}^{-t}\,w^n( t a)\,{\rm d} t\,.\end{eqnarray} 
The second integral on the first line can be
estimated as follows 
\begin{equation}\label{circle}
{R \over |a|} \int\limits_0^\psi \exp \,\left[-{R \cos (\phi-\psi) \over
|a|}\,\right]\, {\rm d} \phi \le \psi\, {R\over |a|} \exp \,\left[-{R \cos
\psi\over |a|}\,\right]\,,\end{equation}
 because $|w(u)|<1$ and $0 <\phi <\psi$.   Since $\psi<\pi/2$, the
expression (\ref{circle}) vanishes for $R\to\infty$. Taking this limit in
(\ref{contour}) one obtains the equality 
 \begin{equation}\label{anpu}
W_{n}^{(+)}(a)=\int\limits_0^\infty\, {\rm e}^{-t}\,  w^n(ta)\,{\rm d}
t\,,\quad  0<\psi<\pi/2\,. \end{equation}

Let us take now $a$ in the fourth quadrant, {\it i.e.}  $-\pi/2<\psi<0$.
A  representation of the form (\ref{anpu}) can be obtained by making a rotation
of the integration contour into  the fourth quadrant up to the negative angle
$\psi$. But this rotation is not allowed for the contour ${\cal C}_+$, since
one encounters the branch cut of the integrand $ w^n(u)$. By Cauchy theorem it
is however easy to relate   $W_n^{(+)}$ to the function $W_n^{(-)}$, defined
by an integral along the contour ${\cal C}_-$. Taking into account the
discontinuity of the function $w(u)$ defined in (\ref{w1}), we have
\begin{equation}\label{cauchy} W_n^{(+)}(a)= W_n^{(-)}(a) +{2\,i\over a}
\int\limits_2^\infty {\rm e}^{-u/a} \,{\rm Im} [w^n(u)]\, {\rm
d}u,\end{equation} where
\begin{equation}\label{imag}
{\rm Im} [w^n(u)]= \left({4+u\over
3\,u}\right)^n\,\sum\limits_{p=0}^{[{n-1\over 2}]} \, (-1)^p C_n^{2p+1} \, 
\left[{\sqrt{(1+u) (u/2-1)}\over 1+u/4}\right]^{2p+1}\,, 
\end{equation}
is the imaginary part for $u>2$  when $u$ approaches the cut from above. It
is convenient to make the change of variable $u=2+y$ in the
last integral of (\ref{cauchy}), writing it as
\begin{equation}\label{y}
2\,i {{\rm e}^{-{2\over a}} \over a} \int\limits_0^\infty {\rm
e}^{-y/ a} \,f_n(y)\, {\rm d} y\,,\end{equation}
where we denoted 
 \begin{equation}\label{fn}
f_n (y)=\left({1+y/6\over 1+y/2}\right)^n\,\sum\limits_{p=0}^{[{n-1\over 2}]}
\, (-1)^p C_n^{2p+1} \,  \left[{\sqrt{(3+y) y/2}\over
3/2+y/4}\right]^{2p+1}\,,   \end{equation}
From (\ref{imag}) it follows that the $f_n(y)$ are algebraic
functions, with poles and branch points for
$y\le 0$ and cuts which can be taken along $y <0$, so the integral (\ref{y}) is
defined. Moreover, we can rotate the integration axis to the lower quadrant,
up to the angle $\psi<0$. This rotation  is allowed also for the function
$W_n^{(-)}$,  leading to a representation identical with the
r.h.s. of (\ref{anpu}). Collecting all the terms we obtain the expression 
\begin{equation}\label{anpl} W_{n}^{(+)}(a)=\int\limits_0^\infty\, {\rm
e}^{-t}\,  w^n(t a)\,{\rm d} t\,+ {2 \rm i}\,{\rm e}^{-{2\over a}}
\int\limits_0^\infty\, {\rm e}^{-t}\,  f_n( t a)\,{\rm d} t\,,\quad
-\pi/2<\psi<0\,. \end{equation} We can use similar arguments for the functions
$W_n^{(-)}$: the contour ${\cal C}_-$ can be rotated towards the fourth
quadrant for negative $\psi$, giving \begin{equation}\label{anml}
W_{n}^{(-)}(a)=\int\limits_0^\infty\, {\rm e}^{-t}\, 
w^n( t a)\,{\rm d} t\,,\quad -\pi/2<\psi<0\,,
\end{equation}
while for $\psi>0$  one must first cross the real axis, picking
up the contribution of the discontinuity of the integrand. This gives 
 \begin{equation}\label{anmu}
W_{n}^{(-)}(a)=\int\limits_0^\infty\, {\rm e}^{-t}\,  w^n(ta) \,{\rm d}
t\, - {2 \rm i}\,{\rm e}^{-{2\over a}} \int\limits_0^\infty\, {\rm e}^{-t}\, 
f_n( t a)\,{\rm d} t\,,\quad 0<\psi<\pi/2\,. \end{equation}
For the PV prescription defined in (\ref{pv}), we obtain from
Eqs. (\ref{anpu}) and (\ref{anmu}): \begin{equation}\label{anpvu}
W_{n}^{(PV)}(a)=\int\limits_0^\infty\, {\rm e}^{-t}\,  w^n( t a)\,{\rm
d} t\, - i\,{\rm e}^{-{2\over a}}\, \int\limits_0^\infty\, {\rm
e}^{-t}\,  f_n( t\, a)\,{\rm d} t\,, \quad  0<\psi<\pi/2\,,
\end{equation}
and from Eqs. (\ref{anpl}) and (\ref{anml}):
\begin{equation}\label{anpvl}
W_{n}^{(PV)}(a)=\int\limits_0^\infty\, {\rm e}^{-t}\,  w^n( t a)\,{\rm d}
t\, + i\,{\rm e}^{-{2\over a}}\, \int\limits_0^\infty\, {\rm
e}^{-t}\,  f_n( t a)\,{\rm d} t\,, \quad  -\pi/2<\psi<0\,.
\end{equation}

The expressions (\ref{anpu}) and  (\ref{anpl}) - (\ref{anpvl})
define holomorphic functions in the right half plane ${\rm Re} a >0$, outside
the real positive axis. When the coupling approaches this
axis from above, {\em i.e.}, for  $a+i \epsilon$, with $a>0$ and
$\epsilon >0$, the expression (\ref{anpvu}) becomes
\begin{equation}\label{anpvur} W_{n}^{(PV)}(a
+i \epsilon)=\int\limits_0^\infty\, {\rm e}^{-t}\,  {\rm Re }[w^n( t\,
a)]\,{\rm d} t\,, \end{equation} since, by
 the definition (\ref{fn}), the imaginary part of the
first integral is exactly cancelled by the last term in  (\ref{anpvu}).
Similarly, for  $a-i \epsilon$  the expression (\ref{anpvl}) gives
   \begin{equation}\label{anpvlr}
W_{n}^{(PV)}(a -i \epsilon)=\int\limits_0^\infty\, {\rm e}^{-t}\,  {\rm
Re }[w^n( t a)]\,{\rm d} t\,, \end{equation} where we used the reality
condition $w(u)= w^*(u^*)$ obvious from (\ref{w1}). Thus, the
functions $W_{n}^{(PV)}(a)$ have no
discontinuity along the positive axis, where they take real values.
 
The expressions   (\ref{anpvu}) and  (\ref{anpvl}) can be analytically
continued from the right half plane to the left one,  ${\rm Re}
a <0$, up to the negative real axis, since the integrals converge and the
integrands have no singularities along the integration contours.
When $a$ approaches the negative real axis,  the imaginary
part of the first term in Eq. (\ref{anpvu}) (or (\ref{anpvl})) is given by the
cut of $w(u)$ along the negative axis, and it is no longer cancelled by 
the imaginary part of the last term, given by the expression (\ref{fn})
analytically continued to negative values of the argument. So,  the functions
$W_{n}^{(PV)}(a)$ have branch cuts for $a<0$. The real analyticity of the
function $w(u)$ implies however the equality 
\begin{equation}\label{refl}
W_{n}^{(PV)}(a+i\epsilon ) = (W_{n}^{(PV)}(a-i\epsilon ))^*.
\end{equation}  
Therefore,  the  $W_{n}^{(PV)}(a)$ are real
analytic functions in the whole complex $a$-plane, except for a cut along the
real negative axis and an essential singularity at $a=0$, seen explicitly in
the representations (\ref{anpvu}) and (\ref{anpvl}). 

For other prescriptions the reality conditions written
above are  not satisfied.  Thus, approaching the positive
real axis from above and from below in (\ref{anpu}) and
(\ref{anpl}), respectively, we obtain \begin{equation}\label{anpur}
W_{n}^{(+)}(a\pm i\epsilon)=\int\limits_0^\infty\, {\rm e}^{-t}\,\{ {\rm Re}[
w^n( t a)] + i \,{\rm Im}[w^n( t a)]\}\,{\rm d} t\,,
\end{equation} with ${\rm Im}[w^n(t)]$ defined in (\ref{imag}). The
functions $W_{n}^{(+)}(a)$ have no discontinuity along the positive axis,
but they are complex for positive values of the coupling. They can also be 
extended into the whole complex plane, except for the essential singularity at
$a=0$ and a branch cut along the negative axis.

At each finite order of truncation in (\ref{cW}), the expanded function 
\begin{equation}
D^{[N]}(s,a) = \sum_{n=0}^{N} c_{n} W_{n}(a) 
\label{cWN}
\end{equation}
will have the same analyticity properties as the $W_n(a)$.  Thus, in
the PV prescription, the $D^{[N]}(s, a)$ have a branch cut along the negative real axis
and an essential singularity at the origin $a=0$. It is instructive to see
what are the consequences for the momentum plane analyticity,  taking the
renormalization point $\mu^2=-s>0$ and the one loop expression of the running
coupling $a(s)= 1/ \ln (-s/\Lambda^2)$. With the usual definition of the
logarithm on the first sheet ($\ln (-s)>0$ for $s<0$), the positive real axis
$a>0$ corresponds to the  part  $s<-\Lambda^2$ of the space-like axis in the
$s$-plane. We obtained no branch cuts in this region, in agreement with the
requirements of unitarity and causality for QCD \cite{Oehme}.  On the other
hand, the negative real axis $a<0$ corresponds to the Landau  region
$-\Lambda^2<s<0$. The $W_n(a)$ have a branch cut along this region, obtained
explicitly by analytic continuation using the representations (\ref{anpvu})
and (\ref{anpvl}). So, in finite orders, the new expansion (\ref{cW}),
calculated outside the Landau region, can be analytically continued inside the
Landau region, where it exhibits a branch cut. Similar properties were
obtained in the large $\beta_0$ limit in \cite{CaNe}.


\begin{figure}\centering
\epsfig{file=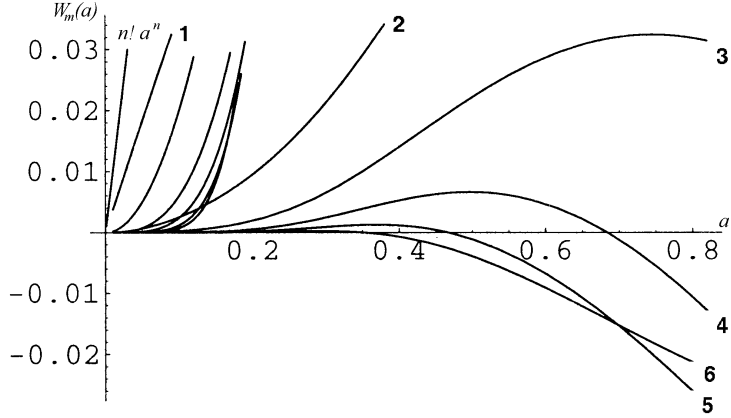,width=14cm}
\caption{ The functions $W_{m}(a)$ defined with the PV
prescription (\ref{pv}),   $m=1, \ldots 6$ for $a$ from $0$ to $0.8$.  
Unlabelled are the perturbative expansion functions $n! a^{n}$ for $n =
1,\ldots, 6$.} \label{fig:4} \end{figure}	

\subsection{Asymptotic expansion of $W_{n}(a)$ for small $a$  }
\label{sec:asym}

In this subsection we  investigate the perturbative expansion of the
functions $W_n(a)$ in powers of $a$.
 As shown in the previous subsection,
$W_{n}(a)$ have singularities at $a=0$, so  their Taylor
expansions around the origin will be  divergent series. We take first $a$
real and positive. The asymptotic  expansion is obtained by applying  Watson's
lemma \cite{Jeff}. Specifically, we consider 
the Taylor expansion \begin{equation}\label{wnser}
w^n(u)=\sum\limits_{k=n}^\infty \xi_k^{(n)} u^k\,,
\end{equation}
which is convergent for $|u|<1$. The sum begins with $k=n$ since, as follows
from (\ref{w1}),  the derivatives $(w^{n})^{(k)}(0)$ vanish for $k<n$ (in
particular $\xi^{(n)}_n=(3/8)^n$).   Choosing a positive number $X<1$, we can
express $w^n(u)$ for $0\leq u\leq X$ as \begin{equation}\label{wnsertr}
w^n(u)=\sum\limits_{k=n}^N \xi_k^{(n)} u^k\,+ R_N^{(n)}\,,
\end{equation}
with a bounded remainder 
 \begin{equation}\label{RN}
|R_N^{(n)}|<   M_n u^{N+1}\,.
\end{equation}
We write now  $W_n(a)$ as \begin{equation}\label{X}
W_{n}(a)={1\over a}\int\limits_0^X\, {\rm e}^{-u/a}\, 
w^n(u) \,{\rm d}u\,+ {1\over a}\int\limits_X^\infty\, {\rm
e}^{-u/a}\,  w^n(u) \,{\rm d}u\,,
\end{equation}
and  insert in the first term the expansion (\ref{wnsertr}). This gives 
\begin{equation}\label{X1}
W_{n}(a)={1\over a}\int\limits_0^X\, {\rm e}^{-u/a}\, 
\,\sum\limits_{k=n}^N \xi_k^{(n)} u^k {\rm d}u\,+ {M_n\over a}\int\limits_0^X\,
{\rm e}^{-u/a}\, \theta_n  u^{N+1} {\rm d}u\,+ {1\over
a}\int\limits_X^\infty\, {\rm e}^{-u/a}\,  w^n(u)\,{\rm d}u\,,
\end{equation}
where $|\theta_n|<1$. Since $|w^n(u)|<1$, the last term in
(\ref{X1}) is bounded as 
\begin{equation}\label{X2}
\bigg| 1/a \int\limits_X^\infty\, {\rm e}^{-u/a}\,  w^n(u)\,{\rm d}u\, \bigg|<
{\rm e}^{-X/a} \,.\end{equation} 
On the other hand, for fixed $k>0$ we have
\begin{equation}\label{Xk}
{1\over a}\int\limits_0^X\, {\rm e}^{-u/a}\,u^k {\rm d}u\,= k! a^k -{1\over
a}\int\limits_X^\infty\,  {\rm e}^{-u/a}\,u^k {\rm d}u\,. \end{equation}
In the last term we make the change of variable $u=X(1+y)$ and use the
inequality $ (1+y)^k \le {\rm e}^{k\,y}$ for $k>0$ \cite{Jeff}, which  
implies that the following estimates \begin{eqnarray}\label{Xk1} {1\over
a}\int\limits_X^\infty\,  {\rm e}^{-u/a}\,u^k {\rm d}u\,= {X^{k+1}\over
a} {\rm e}^{-X/a} \int\limits_0^\infty\,  {\rm e}^{-X y/a}\,(1+y)^k 
{\rm d}y\,\leq   \nonumber\\
 {X^{k+1}\over a} {\rm e}^{-X/a} \int\limits_0^\infty\, {\rm e}^{-y(X/a-k)}
{\rm d}y = {X^{k+1}\over a (X/a-k)}  {\rm e}^{-X/a} \,,
\end{eqnarray}   
are valid for fixed $k$ and small $a$. Therefore Eq. (\ref{Xk}) can be written
as  \begin{equation}\label{Xk2} {1\over a}\int\limits_0^X\, {\rm e}^{-u/a}\,u^k
{\rm d}u\,= k! a^k\,+O \left({\rm e}^{-{X\over a}}\right)  \,.\end{equation} 
By using this estimate in (\ref{X1}) and taking into account the inequality
(\ref{X2}), we  obtain   \begin{equation}\label{Was}
W_{n}(a) = \sum\limits_{k=n}^N \xi_k^{(n)} k! a^k \,+ {\tilde M}_n \,(N+1)!
\,a^{N+1}+  O\left({\rm e}^{-{X\over a}}\right)\,,
\end{equation} 
where ${\tilde M}_n$ is independent of $N$. One has therefore the asymptotic
series   \begin{equation}\label{Was0}
W_{n}(a) \sim \sum\limits_{k=n}^\infty \xi_k^{(n)} k! a^k \,,\quad a\to 0_+\,\,.
\end{equation} 
It is easy to extend the above arguments for complex values of $a$ in the
right half-plane, $|\psi|\leq \pi/2-\delta$ with $\delta>0$.   

\begin{figure}\centering
\epsfig{file=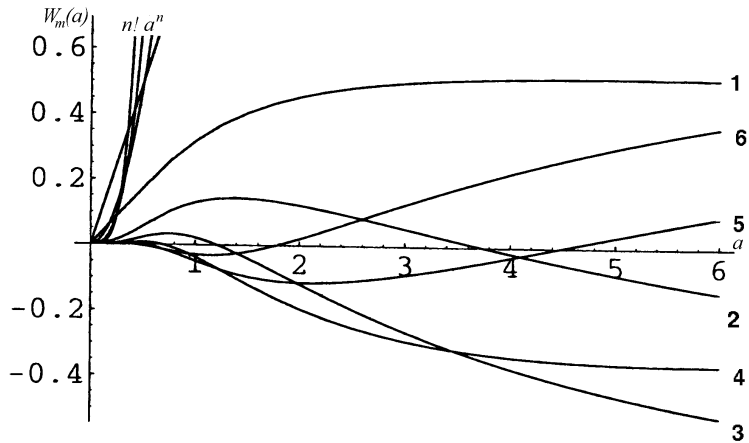,width=14cm}
\caption{The same graph as in Fig. \ref{fig:4}, with $a$ in the interval
$(0,\, 6)$.} \label{fig:5}
\end{figure}
The expansion (\ref{Was0}) is independent of the prescription
required in the definition of $W_n$. 	
(It is formally obtained by inserting
the expansion (\ref{wnser}) in Eq. (\ref{Wn}) and integrating term by term
from 0 to $\infty$.) We notice that the first term of each $W_{n}(a)$ is
proportional to $n! a^n$ with a positive coefficient,  thereby  retaining a
fundamental property of perturbation theory. But the series (\ref{Was0}) 
are divergent: indeed, since the expansions (\ref{wnser}) have their convergence
radii equal to 1, then for any $R > 1$ there are infinitely many $k$ such that  
$|\xi^{(n)}_k| > R^{-k}$ \cite{Jeff}. Actually, the divergence of the series 
(\ref{Was0}) is not surprising, in view of the analyticity 
properties derived in the previous section. 

 For illustration we
give below the expansions of the first $W_n(a)$ \begin{eqnarray} W_{1}(a)
&\sim& 0.375 a - 0.187 a^{2} + 0.457 a^{3} - 1.08 a^{4} + 4.32 a^5+\ldots,
\nonumber \\ W_{2}(a) &\sim& 0.281 a^{2}- 0.422 a^{3} + 1.58 a^{4} - 5.80
a^{5} + 29.78 a^6+\ldots, \nonumber \\  W_{3}(a)&\sim& 0.316 a^{3}- 0.949
a^{4}+ 5.04 a^{5} - 25.95 a^6 +167.99 a^7+\ldots  \label{sim}\,. \end{eqnarray}
The higher powers of $a$  become quickly important in (\ref{sim}),
the expansion coefficients eventually adopting factorial growth.
For instance, the 5th-order coefficients in (\ref{sim}) all equal 5
approximately, while the 10th-order ones are between $5 \times 10^4$
and $9 \times 10^4$, with alternating signs.  
The functions  $W_{n}(a)$ have  divergent perturbative expansions,
resembling the expanded QCD correlation function $D$.

Although the series (\ref{sim}) are divergent, the
functions $W_{n}(a)$ are well-defined (once a prescription has been adopted),
and bounded in the right half plane  ${\rm Re} a > 0$: \begin{equation}
|W_{n}(a)| \leq \frac{ 1}{|a| } \int_{0}^{\infty}{\rm e}^{-u {\rm Re } a/
|a|^2}|w^n(u)|{\rm d}u  <  {|a| \over {\rm Re} a} \, , \label{boundW}
\end{equation} since $|w^n(u)| <1$. For $a$  real and positive the right
hand side of (\ref{boundW}) is equal to unity.  We indicate
in  Figs. \ref{fig:4} and \ref{fig:5} the shape of the first functions
$W_n$, calculated with the PV prescription,  for real values of
$a$.

\subsection{Large-order behaviour of $W_{n}(a)$ and  convergence
conditions } \label{sec:large}

The convergence of the series (\ref{cW})  is not a priori
obvious. Indeed, the expansion 
(\ref{Bw}) of the Borel transform in powers of $w$ converges for points $|w|<
1$ arbitrarily close to the integration axis, but not  necessarily on the
boundary. Intuitively, one may expect that the convergence depends on the
strength of the singularities of the Borel transform.

 We investigated the convergence problem in \cite{CaFi1}, by estimating the
behaviour of $W_{n}(a)$ for large $n$ with the method of 
steepest descent \cite{Zinn}, \cite{Jeff}.
The function $W_n^{(+)}(a)$ defined in (\ref{pm}) can be written as 
\begin{equation}\label{Wp}
W^{(\pm)}_n(a)=\int\limits_{{\cal C}_\pm} {\rm e}^{-F_n(u)}\,{\rm d} u \,,
\end{equation}
  where
\begin{equation}\label{Fn}
 F_n(u)={u\over a}- n \ln w(u)\,.\end{equation}
 The saddle points of the integrand are given by  the  equation
\begin{equation}\label{eq}
{w'(u)\over w(u)}={1\over a n}\,,
\end{equation} 
which has four complex solutions. The technique applied in  \cite{CaFi1}
consists in rotating  the integration axis in the complex
$u$-plane,  without crossing singularities, up to the nearest saddle points
(located in the first (fourth) quadrant for ${\cal C}_+$ ( ${\cal C}_-$ ),
respectively). Omitting the  details given in \cite{CaFi1}, we quote  the
asymptotic  behaviour of $W^{(+)}_n(a)$ for $n \rightarrow \infty$   
\begin{equation}\label{Wnpsaddle}
W^{(+)}_n(a)\approx n^{{1\over 4}} \zeta^n  {\rm e}^{-2^{3/4} (1+i)
{(n/a)^{1/2}}}\,,\quad n\to \infty\,, \end{equation} with $\zeta$ defined
below (\ref{uw}). 
Similarly, the large $n$ behaviour of $W^{(-)}_n(a)$ is
\begin{equation}\label{Wnmsaddle}
W^{(-)}_n(a)\approx n^{{1\over 4}} (\zeta^*)^n
 {\rm e}^{-2^{3/4} (1-i) {(n/a)^{1/2}}}\,,\quad n\to \infty\,.
\end{equation}
These expressions  are valid for complex $a$,   $a=|a|
{\rm e}^{{\rm i}\psi}$, where $\psi$ satisfies the inequality \cite{CaFi1} 
\begin{equation}
|\psi|<\pi/6\, .
\label{psi}
\end{equation}
The convergence of the expansion (\ref{cW}) depends on the ratio
\begin{equation}\label{ratio}
\bigg\vert{c_n W^{(\pm)}_n(a)\over c_{n-1} W^{(\pm)}_{n-1}(a)}\bigg\vert\,.
\end{equation}
As discussed in \cite{CaFi}, if the coefficients $c_n$ satisfy the condition
 \begin{equation}\label{cnb} |c_n| < C {\rm e}^{\epsilon
n^{1/2}}\, \end{equation}
for any $\epsilon >0$,
then the expansion (\ref{cW}) converges for $a$ complex in the domain
\begin{equation}\label{domain}
{\rm Re}[(1\pm i) a^{-1/2}]>0\,,
\end{equation}
which is equivalent to $ |\psi|\leq \pi/2-\delta$.
Since the condition   (\ref{psi}) is more restrictive, it follows that, if the condition
(\ref{cnb}) is satisfied,  
the series (\ref{cW}) converges in the sector defined by (\ref{psi}).

 The coefficients
$c_n$ are obtained by inserting into the Taylor series (\ref{B}) the
expansions in powers of $w$ of the function $u(w)$ defined in (\ref{uw}). A
precise estimate of the behaviour of the $c_n$ starting from a general  form 
of the standard perturbative coefficients  $D_n$ is difficult to obtain. 
In the special case of a Borel transform with  branch point
singularities, considered in \cite{CaFi1}, one can derive the generic
behaviour  \begin{equation}\label{cnb1} |c_n| \leq  C' n^\xi  = C' {\rm
e}^{\xi \ln n}\,,\quad \xi>0\,, \end{equation} which satisfies the convergence
condition (\ref{cnb}). Whether this bound is valid or not in general in QCD is
an open problem, but  (\ref{cnb1}) nevertheless represents a rather
conservative assumption.

In subsection \ref{sec:analyt} we established analyticity properties for the
functions $W_n(a)$ in the complex $a$ plane. The expanded function $D(s,a)$
will be  holomorphic in the domain where the series (\ref{cW}) converges
uniformly. Using the results of subsection \ref{sec:analyt}, we
conclude that $D(s, a)$ is expected to be holomorphic inside  the sector
(\ref{psi}) of the complex $a$-plane, with an essential singularity at the
origin. This region is larger than the horn shaped domain found in
\cite{tHooft} and \cite{Khuri1}. The difference can be easily understood:
indeed, the multiparticle singularities of the correlators in the momentum
plane for  $s\to \infty$, essential for the horn shaped boundaries preventing
the analytic continuation near $a=0$ \cite{tHooft}, \cite{Khuri1}, cannot
appear  in perturbative massless QCD, which is the frame adopted here.

\section{Concluding remarks}\label{sec:concl}

In the present paper we used the analytic continuation in the Borel plane to
define a modified perturbative expansion in QCD, the $n$-th order coefficient
$c_n$ being calculable from the standard perturbative
coefficients $D_k$, $k\leq n $. The new  expansion functions  $W_{n}(a)$
replacing the powers of the coupling constant are defined in (\ref{Wn}) as the 
Laplace transforms of the $n$-th power of the  function $w(u)$, which  
performs  the conformal mapping of the whole doubly-cut Borel plane onto 
a unit disk. 

The definition of the integral
(\ref{Wn}) is ambiguous due to the branch cut $u \geq 2$ along the integration
path. We propose to define $W_{n}(a)$ by choosing the same prescription
as that used to define the expanded function $D(s)$ itself, (\ref{Laplace}),
so as to make the $W_{n}(a)$ resemble $D(s)$ as much as possible. 
The form of the $W_{n}(a)$ is not 
accidental or chosen ad hoc: it is imposed by the analyticity properties 
of the Borel transform, which in turn are determined by the information 
contained in the perturbative coefficients of the QCD correlation 
functions at all orders. 

 We proved that 
 
/i/ the functions $W_{n}(a)$ are analytic in the complex
$a$-plane, with essential and branch point  singularities at $a=0$ 
(subsection 3.2), 

/ii/ their expansions in powers of the coupling constant $a$ are asymptotic divergent
series, the lowest order term of  $W_{n}(a)$ being  proportional to $a^n$
(subsection 3.3), and  		

/iii/ the expansion (\ref{cW}) of the correlation
function in terms of the $W_{n}(a)$ set converges under     
plausible conditions, such that are expected to be satisfied in perturbative QCD 
(subsection 3.4, see also our previous paper \cite{CaFi1}).  By this we avoid 
the fatal divergence of the standard perturbative expansion:  for the new
expansion in terms of the $W_{n}(a)$, the addition of  higher-order terms does
not damage the result, as is the case with the series in powers of $a$.   The
convergence property is the main merit of the new expansion. 

 The prescription required  for calculating the Laplace integral in the
definition (\ref{Wn}) reflects  the intrinsic ambiguity of perturbation theory,
which   originates from the infrared regions of the Feynman diagrams and is
manifest in the presence of the IR renormalon cut.   Once a definition of the
perturbative ambiguity is adopted (as the difference between two integration
prescriptions, for instance), the series (\ref{cW}) provides a
systematic calculation of this ambiguity at higher orders, since  the expansion
 in terms of the $W_{n}(a)$ is convergent. This is to be contrasted with
the standard expansion, where the ambiguity is defined in a less precise way by
truncating the series at some finite order, beyond which the terms start to
increase. Unlike in this procedure,    the effects of
the UV and IR parts of the Feynman diagrams have been disentangled 
in our approach. 

Thus,
using the analytic continuation in the Borel plane, we have been 
able to separate two problems that are usually interconnected: the 
divergence of the series (which can be solved within perturbation theory),
and the problem of the intrinsic infrared  ambiguity of perturbation
theory. This ambiguity, expressed in the prescription dependence of the
correlator and the expansion functions $W_{n}(a)$, can be removed only when
nonperturbative effects are included.

\vskip1cm 
{\bf Acknowledgements:}  The authors thank Prof. S. Randjbar-Daemi and the High
Energy Section of the Abdus Salam International Centre for Theoretical
Physics in Trieste for hospitality. One author (J.F.) is indebted to 
Prof. G. Altarelli for hospitality at the CERN Theory Division. Interesting
discussions with Dr. Ivo Vrko\v{c} on the mathematical aspects of this work (in
particular his advice on the methods used in subsection 3.2) are
gratefully acknowledged. This paper was partially supported by the Romanian 
Academy, under the Grant 49/2000, and by the Ministry of Industry 
and Trade of the Czech Republic, project RP-4210/69.


\begin{thebibliography}{99}


\bibitem {tHooft}
G. 't Hooft, in: {\it The Whys of Subnuclear Physics}, Proceedings of
the 15th International School on Subnuclear Physics, Erice, Sicily,
1977, edited by A.~Zichichi (Plenum Press, New York, 1979), p.~943.

\bibitem{Khuri1} N. N. Khuri, Phys. Rev. D{\bf 23}, 2285 (1981).

\bibitem{Zakh} V. Zakharov, Nucl.Phys. B{\bf 385}, 452 (1992).

\bibitem{Muel1} A.H. Mueller, in {\em QCD - Twenty Years Later}, Aachen 1992, 
edited by P. Zerwas and H. A. Kastrup (World Scientific, Singapore, 1992).

\bibitem {BeZa} M. Beneke and V. I. Zakharov, Phys. Lett. B{\bf 312}
340 (1993).

\bibitem{Grun} G. Grunberg, Phys. Lett. B {\bf 304}, 183 (1993).

\bibitem {Bene} M. Beneke, Phys.\ Lett.\ B {\bf 307}, 154 (1993); 
Nucl.\ Phys.\ B {\bf 405}, 424 (1993).

\bibitem {Broa} D. Broadhurst, Z.\ Phys.\ C {\bf 58}, 339 (1993).

\bibitem{Oehme}  R. Oehme, $\pi N$ Newsletters, {\bf 7}, 1 (1992); Int. J.
Mod. Phys. A {\bf 10} 1995 (1995),


\bibitem{Milt} K.A. Milton and I.L. Solovtsov, Phys.Rev. D {\bf 55}, 5925
(1997). 

\bibitem{Shir} D.V. Shirkov, Theor.Math.Phys.{\bf 127} 409-423 (2001). 

\bibitem{Stev} P. M. Stevenson, Phys. Rev. D{\bf 23}, 2916 (1981).

\bibitem{Grun1} G. Grunberg,  Phys. Lett. B{\bf 95}, 70 (1980).

\bibitem{Max} C. J. Maxwell, Phys.Lett. B{\bf 409}, 450 (1997).
  
\bibitem{SaEl} M. A. Samuel, J. Ellis and M. Karliner, Phys. Rev. Lett. {\bf
74}, 4389 (1995).

\bibitem{Ellis} J. Ellis, M. Karliner and M.A. Samuel, Phys. Lett. B{\bf 400},
197 (1997).

\bibitem{Jent} U.D. Jentschura, J. Becher, E. J. Weniger, G. Soff, Phys. Rev.
Lett. {\bf 85}, 2446 (2000). 



\bibitem {BBB} P. Ball, M. Beneke and V.M. Braun,  Nucl. Phys. B{\bf 452},
563 (1995).

\bibitem {Neub} M. Neubert, Phys.\ Rev.\ D{\bf 51}, 5924 (1995).


\bibitem{LoMa} C. N. Lovett-Turner and C. J. Maxwell,  Nucl. Phys. B{\bf 452},
188 (1995).

\bibitem{Bene1} M. Beneke, Phys. Rep. {\bf 317}, 1-142 (1999).

\bibitem{BeBr} M. Beneke and V.M. Braun, hep-ph/0010208, Boris Ioffe
Festschrift "At the Frontier of Particle Physics/Handbook of QCD", edited by M.
Shifman (World Scientific, Singapore, 2001).
 

\bibitem{Hardy} G.N. Hardy, {\em Divergent Series} (Oxford University Press, 
New York, 1949).

\bibitem {Fischer} J. Fischer, Fortsch. Phys. {\bf 42}, 665 (1994);
 Int. J.Mod.Phys. A {\bf 12}, 3625 (1997).

\bibitem{Muel} A. Mueller, Nucl.Phys. B {\bf 250}, 327 (1985).

 \bibitem {Alta}  G. Altarelli, P. Nason and G. Ridolfi, Z. Phys. C {\bf 68},
257 (1995).   

\bibitem {SoSu} D.E. Soper and L. R. Surguladze, Phys. Rev. D {\bf 54}, 
4566 (1996). 
 


\bibitem{CaFi} I. Caprini and J. Fischer, Phys.Rev D{\bf 60}, 054014 
(1999).

 \bibitem{CaFi1} I. Caprini and J. Fischer, Phys.Rev D{\bf 62}, 054007 (2000). 
 
\bibitem{Jent1} U.D. Jentschura, E. J. Weniger, G. Soff, J. Phys. G26, 1545
(2000).

\bibitem{CvLe} G. Cveti\v c and T. Lee, Phys. Rev D{\bf 54}, 014030 (2001).



\bibitem{BBK} M. Beneke, V.M. Braun and N. Kivel, Phys. Lett. B {\bf 404}, 
315 (1997).

\bibitem{Lipa} L.N.Lipatov, Sov.Phys. JETP {\bf 45}, 216 (1977).


\bibitem{CiFi} S. Ciulli and J. Fischer, Nucl. Phys. {\bf 24}, 465 (1961).


\bibitem{CCF} I. Ciulli, S. Ciulli and J. Fischer, Nuovo Cim. {\bf 23}, 1129
(1962).


\bibitem{Zinn} J. Zinn-Justin, {\em Quantum Field Theory and Critical 
Phenomena}  2nd edition, (Oxford University Press, 1995), p. 923 - 928. 

\bibitem{CaNe} I. Caprini and M. Neubert, JHEP {\bf 03}, 007 (1999). 


\bibitem{Jeff} H. Jeffreys, {\em  Asymptotic Approximations} (Clarendon Press, 
Oxford, 1962).


\end{thebibliography}
\end{document}